\documentstyle[12pt,aaspp4]{article}
\begin{document}
\title{EGRET Observations of Gamma Rays \\ from Point Sources with
Galactic Latitude \\ $\bf +10\arcdeg < b < +40\arcdeg$
  }
\author{P. L. Nolan\altaffilmark{1,10},
D. L. Bertsch\altaffilmark{2},
J. Chiang\altaffilmark{3},
B. L. Dingus\altaffilmark{2,8},
J. A. Esposito\altaffilmark{2,8},
C. E. Fichtel\altaffilmark{2},
J.~M.~Fierro\altaffilmark{1},
R. C. Hartman\altaffilmark{2},
S. D. Hunter\altaffilmark{2},
G. Kanbach\altaffilmark{4},
D. A. Kniffen\altaffilmark{5},
Y.~C.~Lin\altaffilmark{1},
J.~R.~Mattox\altaffilmark{6,8},
H.~A. Mayer-Hasselwander\altaffilmark{4},
P.~F. Michelson\altaffilmark{1},
C. von~Montigny\altaffilmark{2,9},
R.~Mukherjee\altaffilmark{2,8}
E.~Schneid\altaffilmark{7},
P.~Sreekumar\altaffilmark{2,8},
D. J. Thompson\altaffilmark{2},
T. D. Willis\altaffilmark{1}
}
\altaffiltext{1}{W. W. Hansen Experimental Physics Laboratory,
Stanford University, Stanford, CA 94305-4085}
\altaffiltext{2}{NASA/Goddard Space Flight Center, Code 662,
Greenbelt MD 20771}
\altaffiltext{3}{CITA, University of Toronto, 60 St George Street, Toronto, 
 Canada M5S 1A1}
\altaffiltext{4}{Max-Planck Institut f\"ur Extraterrestrische Physik,
Giessenbachstra\ss e, 85748 Garching bei M\"unchen, Germany}
\altaffiltext{5}{Hampden-Sydney College, P. O. Box 862, Hampden-Sydney
VA 23943}
\altaffiltext{6}{Astronomy Department, University of Maryland, College Park
MD 20742}
\altaffiltext{7}{Northrop Grumman Corporation, Mail Stop A01-26,
Bethpage NY 11714}
\altaffiltext{8}{Universities Space Research Association, Code 610.3,
NASA/GSFC}
\altaffiltext{9}{NAS/NRC Research Associate}
\altaffiltext{10}{pln@egret1.Stanford.EDU}
\authoremail{pln@egret0.stanford.edu}
\date{CGRO/EGRET/SU/PLN/1994/DEC/9}

\begin{abstract}
The EGRET instrument aboard the Compton Gamma Ray Observatory (CGRO) has
completed the first all-sky survey in high-energy gamma rays and
has repeatedly viewed selected portions of the sky.
Analysis of the region with galactic latitude $+10\arcdeg < b <
+40\arcdeg$ indicates the presence of nineteen point sources, including 
nine which can be identified as
active galactic nuclei, some of which have been reported previously, as well
as ten other sources with no definite counterparts.
Using the combined exposures from Phase 1 and Phase 2 of the CGRO
viewing program, the spectra, time variability, and positions of all
detected sources in this region are determined.
It is tentatively suggested that one of the unidentified sources might be
associated with the radio galaxy Centaurus A.
\end{abstract}


\newcommand{\TS}{\mbox{$T_s$}}
\newcommand{\sqTS}{\mbox{$\sqrt{T_s}$}}

\section{Introduction}

The all-sky survey by the Compton Gamma Ray Observatory 
 has led to the detection of a number
of high-energy gamma ray sources with the
Energetic Gamma Ray Experiment Telescope (EGRET).
EGRET covers the high-energy gamma ray range from about 30 MeV
to over 20 GeV.
It produces an image of a circular region of the sky with
maximum sensitivity at the center of the field of view, falling to
half at about 18\arcdeg from the center and to one-sixth at 30\arcdeg.
Individual photons are detected with an angular accuracy which varies
with the photon energy from about 0.5 degrees at the highest energy
to 10 degrees at the lowest.
The properties of the instrument are discussed by Kanbach et al. (1988, 1989)
and by Thompson et al. (1993a).
The full sky survey provided for the first time an opportunity to
study in detail high-energy gamma ray emission from point sources
as well as from extended regions of diffuse emission.
Individual observations, typically one to three weeks long,
provided an opportunity to examine the emission from potential
point sources while adding all the observations during the survey
allowed weak, steady sources to be enhanced.
This analysis covers Phase 1 and Phase 2 of the CGRO observing
plan, from the spacecraft's launch in April 1991 to August 1993.

This paper covers the detected point sources whose galactic latitude
is between +10\arcdeg and +40\arcdeg.
Sixty-five separate observations contributed useful data for this region.
Other portions of the sky are discussed in other papers (Dingus et al. 1995;
Lin et al. 1995b; Sreekumar et al. 1995; Bertsch et al. 1995).
The sky is divided into zones of different galactic latitude for two
reasons.
First, different types of point sources are found in different parts
of the sky.
The pulsars which EGRET has detected are all within a few degrees
of the galactic plane (Thompson et al. 1994), and it is possible that many other
point sources belong to other galactic populations.
Above 10 degrees of latitude, it is likely that the sources are predominantly
extragalactic.
Second, the galactic diffuse emission has a strong influence on the
ability to distinguish point sources.  
Within 10 degrees of the Galactic plane, most point sources are
weak ``bumps'' on a strong, structured background.
At higher latitudes, the diffuse emission is much weaker and more
uniform.
The statistical criteria used in the data analysis are necessarily
different at different latitudes.

The first EGRET catalog (Fichtel et al. 1994a) contains all
the EGRET point sources detected in the Phase 1 observations.
This paper covers many of the same sources, but there are some 
differences.
With more data, positions and fluxes are known more precisely, and some
marginal sources are now detected significantly.
Information on spectra and time variation are included here.
Some variable sources not detectable in Phase 1 flared in Phase 2
and became detectable.
With an improved model of the Galactic diffuse emission, a number of
spurious sources have been removed from the list.
Point sources can now be studied in the zone with $0\arcdeg < \ell < 30\arcdeg$,
$10\arcdeg < b < 20\arcdeg$, which was omitted from the first catalog because
of shortcomings in the old diffuse model. 
The second EGRET catalog (Thompson et al. 1995) will use the same data used 
here, but the point sources are not discussed in the same detail.

This paper will not discuss the Ophiuchus Cloud, an extended gamma-ray source
covered in detail by Hunter et al. (1994).

\section{Observations and Analysis}

The standard EGRET processing operations were applied to the data
(Bertsch {\it et al.} 1989).
Maps were made of the number of photons detected in 0.5\arcdeg pixels 
and the instrument exposure was calculated for each pixel.
Photons out to 30\arcdeg from the center of the instrument
field of view were accepted.
Maps were made for each individual viewing period as well as for
the whole of Phase 1 and 2.
A list of the viewing periods in Phase 1 is contained in Fichtel
et al. (1994a).
The Phase 1 and 2 viewing periods will be listed in the second EGRET catalog
(Thompson et al. 1995).

Source fluxes are derived by a maximum-likelihood analysis 
(Mattox et al. 1996) which
uses the detector point spread function to model any excess above 
a predicted map of the diffuse Galactic emission (Bertsch et al. 1993, Hunter
et al. 1995) and a (presumed extragalactic) isotropic component.
Recent improvements in this prediction have made it possible to resolve
point sources in the region $0\arcdeg < \ell < 30\arcdeg$, 
$10\arcdeg < b < 20\arcdeg$, which formerly had an implausible number of
deviations from the diffuse model (Fichtel et al. 1994a).
The levels of the predicted Galactic and isotropic diffuse emission are allowed
to vary over limited regions in order to allow for local inaccuracies
of the model.
The statistical significance of a detected point source at a fixed
position is estimated from its test statistic $\TS
\equiv 2\log(L_1/L_0)$, where $L_1$ and $L_0$ are the
likelihoods of the models with the source present and absent.
The strengths of the source and the diffuse components are adjusted to 
maximize \TS.
If no real point source is present, \TS\ at any chosen point on the
sky will be a random variable
distributed as $\chi^2(1)$ (Mattox et al. 1996).
Thus \sqTS\ is approximately the significance of a detection,
expressed in Gaussian ``sigmas.''

To search for possible point sources we first examined a map of the
whole observation containing photons with energy $> 100$ MeV.
This choice has proven to be a useful compromise between the better 
angular resolution
that could be obtained by using only higher-energy photons
and the larger number of photons that could be obtained by admitting
lower energies.
A source is accepted for further analysis if $\sqTS > 4$.
Monte-Carlo simulations show that there should be about 2
spurious sources in the whole sky with $\sqTS > 4$ (Mattox et al. 1996),
in agreement with with approximately Gaussian distribution of
\sqTS.
Thus there should be less than one spurious detection in this region.
To look for other sources with unusual spectra we also searched maps
in other energy bands: 30--100 MeV, 100--300 MeV, 300--1000 MeV, and
$>$1000 MeV.
There were no new detections with $\sqTS > 4$.
We also searched the individual viewing period maps and found no
new sources.

Positions for point sources are estimated by mapping the value of \TS.
A trial point source is placed in many different locations; a different
\TS\ value is obtained at each.  
The maximum point is the most likely location of the source.
Confidence regions are chosen to be areas in which \TS\ is greater than
a cutoff value.
The cutoff is equal to the maximum \TS\ less the appropriate critical value of the
$\chi^2(2)$ distribution.
The positions estimated here were derived in two different ways.
First, the single E $>100$ MeV map was examined.  
Second, \TS\ maps for three energy bands (100--300 MeV, 300--1000 MeV,
and $>$1000 MeV) were added.  
The 30--100 MeV band was omitted because of its poor angular resolution in
comparison with the others.
For the objects studied here,
the second procedure usually produces better results because it takes
better advantage of the narrower point spread function at high energy.
The smaller of the two confidence regions is reported here.
There are still a few regions in which there are possible difficulties
in the diffuse model, as shown by large, unsymmetric error regions, 
suspicious variation between energy bands, or concentrations of weak
flux excesses.
Such cases will be noted for individual source detections.

The positions of the detected point sources are compared with a list
of candidate sources.
The extreme AGN portion of this list includes radio-loud ($> 1$ Jy)
quasars, OVV quasars, high-polarization quasars, superluminal
objects, and BL Lac objects (Fichtel et al. 1994a).
In addition a selection of prominent Seyfert galaxies and radio-quiet 
quasars was added.
Many Galactic objects were also included, such as pulsars, X-ray binaries,
globular clusters, X-ray bursters, supernova remnants, and 
COS-B gamma-ray sources.
The total number of candidates is slightly more than 1000,
of which roughly 200 fall between Galactic latitudes +10\arcdeg
and +40\arcdeg.
The probability of even one chance coincidence with an object from this
list is less than 10\% for a typical error circle of radius $<1$ degree.
Approximately half of our detected sources can be identified with
objects from this list, which indicates that most of the identifications
are probably correct.

Spectra of the detected sources can be determined from the fluxes
measured in the different energy bands.
A power law form without breaks is assumed for the spectrum.
A model spectrum is propagated through a model of the detector
response function (Thompson et al. 1993a) and compared with the
detected flux by a $\chi^2$ test.
The normalization and spectral index are adjusted to obtain the
best agreement with the data (Fichtel et al. 1994b).
The uncertainty in the spectral index is based on the size of the
region in parameter space in which $\chi^2 < \chi^2_{min}+2.3$.
This is appropriate if there are two interesting parameters, the 
spectral index and the normalization (Lampton, Margon, \& Bowyer 1976).  
It might be argued that the index
is the only  interesting parameter in this situation.  In that case
the appropriate confidence region would be $\chi^2 < \chi^2_{min}+1$,
so the uncertainty could be decreased by a factor $(2.3)^{-{1 \over 2}}$.
Spectra for all the sources can be determined in the four energy
bands described above.
For strong sources these can be checked by dividing the 30 MeV --
10~GeV range into ten narrow bands.
Empirical corrections are applied to the flux for energy less than
70 MeV (Thompson et al. 1993b).

\section{Results}

The main results for the combined data are found in Tables 1 and
2.  
There were 19 point sources which met the criteria described above.
These are all presented together in the Tables, abandoning the 
distinctions (Fichtel et al. 1994a) between identified and unidentified sources and between
significant detections ($\sqTS > 5$) and marginal ones ($4 <  
\sqTS < 5$).
Each source has been assigned a name based on the celestial (J2000)
coordinates of its most likely position, for instance J0009+73.  
In some cases the numerical portions of these names differ slightly
from those which have been used to identify these sources in previous
publications because of more data and the improved diffuse model.

Table 1 contains information about the flux, position, and identification
of the sources.
The photon flux for energy $> 100$ MeV is presented, along with the
value of \sqTS\ in this band.
The Galactic coordinates of the most likely position are given, as well
as the radius of the 95\% confidence error circle in arcminutes.
For the identified sources, the counterpart is named and its distance
from the most likely position is given.
The redshifts of the counterparts are shown, if known, and the
gamma ray luminosity in the range 100 MeV -- 10 GeV is calculated.
The luminosity calculation assumes isotropic emission,
$q_0 = {1 \over 2}$, and $H_0 = 75$ km s$^{-1}$ Mpc$^{-1}$.
The photon number spectral index $\alpha$ is taken from Table 2.
A $K$ correction factor of $(1+z)^{-(\alpha+2)}$ has been applied so the
luminosities can be compared for the same emitted energy band.

Table 2 contains information about the spectra of the sources in
broad energy bands.
Photon fluxes are presented for the four standard energy bands:
30--100 MeV, 100-300 MeV, 300-1000 MeV, and $>$1000 MeV.
From these values a photon number spectral index is produced, assuming
a spectrum of the form ${dN \over dE} \propto E^\alpha$.
All of the energy spectra were consistent with the simple power law
model according to a $\chi^2$ test.
There is no compelling evidence for a spectral break or cutoff.
Most of the spectra, however, are of such low quality that no meaningful
individual limits on the shape of the spectrum can be derived.
In such cases the spectral index is just a rough guide to the hardness
of the spectrum.
The average energy spectra in 10 narrower energy bands are displayed 
in Figure 1.

To search for possible variability,
the source fluxes were examined in each of the individual viewing
periods.
A fairly stringent criterion was used: a $\chi^2$ test must show less
than 5\% probability that the source flux is constant.
On this basis only five sources showed definite evidence for variation,
J0009+73 (2.6\% probability of not being variable), J0720+21 (1.7\%), 
J0744+54 (2.3\%), J0826+70 (4.5\%), and J1835+59 (1.9\%).
Figure 2 shows the $>$100 MeV light curves for three of these objects.
The other two are not presented here because they have been discussed fully
elsewhere (J0720+71 identified as 0716+714 in von Montigny et al. 1995 and
Lin et al. 1995a; J0744+54 in Mukherjee et al. 1995).
Some other sources have been described as variable (J1626-24 in Hunter
et al. 1994; J0826+70 in Michelson et al. 1994), but they do not pass
our $\chi^2$ test.
It is possible that all of the detected sources may be variable to some 
degree, but the
measurements are not precise enough to detect small changes in the
weaker ones.

In a few cases some sources were bright enough to be detected with 
$\sqTS > 4$ in individual viewing periods.
In these cases, it is possible to derive spectra for the individual
observations.
The resulting fluxes and spectral indices are shown in Table 3.
Note that the five objects listed here are the same ones with
significant variability.
The spectra of two objects, J0721+71 and J1835+59, show significant
variation in spectral index.
There is no obvious correlation between spectral index and flux.
The individual spectra of J1835+59, the brightest and most often
observed source, are shown in Figure 3.

Because of the recent report (Quinn et al. 1995) of the detection
of 0.3 TeV gamma rays from the nearby BL Lac object Markarian 501
with a flux of $8 \times 10^{-12}$ photons cm$^{-2}$ s$^{-1}$,
we searched for emission in the EGRET energy range.
None was detected, and the 95\% confidence upper limit is
$1.0 \times 10^{-7}$ photons cm$^{-2}$ s$^{-1}$ for E $> 100$
MeV.
If the spectrum spanning the two energy bands is assumed to be a
single power law, then it must be flatter than $\rm E^{-2.2}$.

\section{Discussion}

This list cannot be considered a complete, flux-limited survey in a strict
sense.
The limiting flux is a function of the local diffuse emission.
A source could appear on the list either by having steady emission for
the whole survey period or by having a relatively brief flare.
Nevertheless it is possible to say that the minimum average flux detected
is about $10^{-7}$ photons cm$^{-2}$ s$^{-1}$ for energy $> 100$ MeV.

There is marginal evidence for a difference in the spectra of the
identified and unidentified objects.
The mean spectral index of the identified sources is $-2.46$, with a
standard deviation of 0.51; the mean and deviation for the unidentified
sources is $-2.16$ and 0.39.
A Student-t test shows that the hypothesis that the two sets are drawn from
the same parent population can be rejected with only 83\% confidence.
It remains possible that the unidentified sources may not be a homogeneous
class.

\subsection{Identified Sources}

The gamma source counterparts are blazars, that is BL Lac objects
or radio-loud quasars with a flat radio spectrum, high optical polarization,
or rapid variability (Angel \& Stockman 1980; Burbidge \& Hewitt 1992).
Our confidence in these identifications is based on a statistical argument:
there are not many blazars, and many of them fall in the EGRET error circles
(von Montigny et al. 1995).
In these objects, the apparent gamma-ray
luminosity dominates all other wavelength bands if the emission
is assumed to be isotropic.
This gamma-ray brightness is seen as evidence for beaming by a
relativistic jet.
Von Montigny et al. (1995) describe many of the specific models which
have been proposed to explain the gamma-ray emission.

Most of the identified sources in this region have been discussed
in some detail by von Montigny et al. (1995).  
Because of improvements in the model of Galactic diffuse gamma
ray production (Hunter et al. 1995), the significance of some of
these detections has changed.
Also some source positions have shifted enough to affect identifications.
The sources 0829+046 and 1741$-$038 in von Montigny et al. (1995) do not
appear in this paper because their detections no longer pass the
$\sqTS > 4$ test.
Conversely, J0740+18 and J1734$-$13 don't appear in von Montigny et al. (1995), 
but they are in the present list.
Other published possible blazar identifications from the EGRET data are 
noted below.
Unless otherwise noted, redshifts and basic characterizations are taken
from the compilation of Hewitt \& Burbidge (1993) and the NASA Extragalactic
Database (NED) (Helou et al. 1991).
Following are the identified sources.

J0720+71 = 0716+714.
This is a BL Lac object with a flat radio spectrum.
Its redshift is unknown, but it is probably more than 0.3 (Witzel et al. 
1988).
The EGRET observations of this object have also been discussed in detail
by Lin et al. (1995a).
Its gamma ray flux is variable.

J0740+18 = 0735+178.
This is a BL Lac object.
According to Stevens et al. (1994), its high-frequency radio emission 
underwent a smooth decrease during the entire period of the EGRET 
observation after outbursts in 1989 and early 1991.

J0808+49 = 0804+499 = OJ 508.
This is a highly-polarized QSO.
The EGRET observations of this object have also been discussed 
by Thompson et al. (1993b).

J0826+70 = 0836+710 = 4C +71.07.
This is a flat-spectrum QSO.
The EGRET observations of this object have also been discussed 
by Thompson et al. (1993b).
This source is variable with confidence $> 95$\%. 
Its light curve is presented in Figure 2(b).
It is the most luminous of all the identified sources in this section
of the sky.

J0831+24 = 0827+243 = OJ 248.
This is a flat-spectrum QSO.
The EGRET observations of this object have also been discussed 
by Michelson et al. (1994).
There is some uncertainty about its redshift.
We adopt the value 0.941 of Steidel \& Sargent (1991) rather than
2.046 as listed in the NED database.

J1315$-$34 = 1313$-$333.
This is a flat-spectrum QSO.

J1626$-$24 = PKS 1622$-$253.
The EGRET observations of this object have also been discussed in detail 
by Hunter et al. (1994).
This is located behind the Ophiuchus Cloud, a strong, compact, diffuse
gamma ray source.
Although it is not listed in the Hewitt \& Burbidge (1993) catalog,
Impey \& Tapia (1990) identify it as a quasar.
The radio spectrum is bright and flat enough to qualify it as a blazar
candidate.
No redshift value is known.

J1734$-$13 = 1730$-$130.
This is a flat-spectrum, variable QSO.
This is the only source in this list for which a better position was 
obtained by using the E $>100$ MeV band.
It is located in a region where the galactic diffuse region is
strong and difficult to model.

J1738+51 = 1739+522 = 4C +51.37.
This is a flat-spectrum QSO.

\subsection{Unidentified Sources}

Ten point sources are not identified with AGN counterparts.
At such high galactic latitudes, it is probable that most of these
are AGN of types similar to the identified sources.
Some, however, might be unidentified Geminga-like pulsars or other nearby
galactic objects.
However, most of the unidentified sources have steeper spectra than
the EGRET-detected pulsars (Thompson et al. 1994).
Most of the error circles are so large that the NED or SIMBAD databases
will contain several possible counterparts.
Most of those are weak radio sources or IRAS infrared sources, which are
so numerous that several chance coincidences are expected.
We will comment on the brightest gamma sources and the ones with bright
($> 0.25$ Jy) potential radio counterparts.

J0009+73.
This source is variable with confidence $> 95$\%.
The light curve is shown in Figure 2(a).
It is coincident in position with the supernova remnant CTA 1, a plerion
with a central point X-ray source (Seward 1990), $1.2 \times 10^4$ years old
(Sieber, Salter, \& Mayer 1981), at a distance of 1.1 kpc (Milne 1979).
At this distance, the gamma luminosity would be about $10^{34}$ erg
s$^{-1}$. 
This luminosity and the flat spectrum are appropriate for a pulsar, 
but the variability is not.
Sturner \& Dermer (1995) have proposed that several other EGRET sources
are remnants, and they describe several mechanisms by which gamma rays
might be produced.
It is possible for the emission from a remnant to vary in a rew months
(de Jager \& Harding 1992), so this identification can't be ruled out.
The flat-spectrum QSO 0016+731 is also less than a degree away, although 
the position is not formally consistent.

J0445+62.
This might be identified with the extragalactic radio sources 4C +61.11 
(27\arcmin\ away, 2.4 Jy at 178 MHz) or 0437+6139 (29\arcmin\ away, 0.4 Jy
at 1.4 GHz).
 
J0500+59.
This might be identified with the extragalactic radio sources 4C +59.05
(41\arcmin\ away, 2.5 Jy at 178 MHz), 0500+5935 (49\arcmin\ away, 2.6 Jy
at 1.4 GHz), or 0451+5945 (61\arcmin\ away, 0.57 Jy at 1.4 GHz).
The position of this source should perhaps be regarded with a certain
degree of caution.
The error circle is elongated, it is detected in only two of the four 
broad energy bands, and the
two best positions don't agree within their formal uncertainties.
These factors might indicate remaining problems in the diffuse model.

J0744+54. This source is variable (Mukherjee et al. 1995).
It could plausibly be identified with the radio source 0738+5451, 
14\arcmin\ away,
which has a radio flux of 0.27 Jy at 4.85 GHz and a flat spectrum.


J1326$-$43.
This error circle contains no known blazars, but 31\arcmin\ from the
center is Centaurus A = NGC 5128, an unusual radio galaxy with an 
X-ray jet (Feigelson et al. 1981)
and large radio lobes
(Meier et al. 1989) located only 3.5 Mpc away (Hui et al. 1993).
This identification must be tentative with such a large error 
circle.  
Also, the diffuse emission in this region shows some low-level
structure not found in the model.
Such irregularities might disturb the position estimate or the
significance of the detection.
This possible identification has not been mentioned in any previous
list because recent revisions of the galactic diffuse model caused
the source position to shift.
Other catalogued objects in the circle include several normal
stars, several normal galaxies, several IRAS infrared sources, 
the pulsar PSR 1325-43, and the supernova SN 1986G.
None of these shows any characteristics that would make it likely to
be a strong emitter of high-energy gamma rays.
Nice, Sayer, and Taylor (1994) searched part of the error circle for
new pulsars, but found none.

It is plausible that Cen A could be a high-energy gamma ray source.
It is the brightest extragalactic source of hard X-rays near 100
keV and its spectrum doesn't show the sharp high-energy cutoff typical
of Seyfert galaxies (Kinzer et al. 1995).
The EGRET spectrum extrapolates to match fairly well with the 
contemporaneous COMPTEL measurements of Collmar et al. (1993) around 1 MeV.
The spectral index would have to steepen by about 1 between 100 keV and the
EGRET energy range.
This gradual steepening is consistent with a nonthermal 
synchrotron/Compton model of the emission (Grindlay 1975;
Mushotzky et al. 1978; Beall \& Rose 1980), probably
not with the Compton reflection model (Zdziarski et al. 1990;
Skibo, Dermer, \& Kinzer 1994).
The gamma ray luminosity of $8.3 \times 10^{40}$ erg s$^{-1}$
is only a few hundred times our own galaxy's.




J1835+59.
This is the brightest of the EGRET unidentified sources.
It was discussed by Nolan et al. (1994).
Since then the error circle has been refined to a smaller size,
so the list of possible counterparts in the standard catalogs has been 
reduced to one: IRAS F18342+5913.
This is is a faint point source
in the IRAS wavelength bands, and nothing else is known about it.
No radio source in the Green Bank survey (Becker, White, \& Edwards 1991)
is consistent with the error circle.
Nice, Sayer, and Taylor (1994) searched for pulsars in the error circle,
and found none.
This source is variable on a time scale of weeks.
Its light curve is shown in Figure 2(c).
There is also some evidence that the spectral index varies.

\section{Summary}

Nineteen point sources of $> 100$ MeV gamma rays can be identified in
this region of the sky.
Nine of them can be identified with blazars.
This identification can be made confidently because of the relative
rarity of blazars.
Five of the brightest sources, both identified and unidentified,
are variable in flux.
One of the bright unidentified sources is coincident with the plerion
supernova remnant CTA 1, and another is consistent with the position of
the radio galaxy Centaurus A.  
The brightest of the unidentified sources has a 9\arcmin\ radius error
circle containing no prominent identification candidates in other 
wavelength bands.

\acknowledgments
The EGRET team gratefully acknowledges support from the following:
Bundesministerium f\"ur Forschung und Technologie grant 50 QV 9065
(MPE), NASA Grant NAG5-1742 (HSC), NASA Grant NAG5-1605 (SU),
and NASA Contract NAS5-31210 (GAC).
This research has made use of the NASA/IPAC Extragalactic 
Database (NED) which is operated by the Jet Propulsion Laboratory,
California Institute of Technology, under contract with the National
Aeronautics and Space Administration,
and of the Simbad database, operated at CDS, Strasbourg, France.


\begin{deluxetable}{lrrrrrcccc}
\scriptsize
\tablecaption{Gamma Ray Sources \label{table1}}
\tablehead{
\colhead{EGRET\tablenotemark{a}} & \colhead{${\rm F}_{>100}$\tablenotemark{b}} 
& \colhead{\sqTS\tablenotemark{c}} 
& \colhead{$\ell$\tablenotemark{d}}
& \colhead{b\tablenotemark{d}} & \colhead{95\%\tablenotemark{e}} 
& \colhead{ID\tablenotemark{f}} & \colhead{$\Delta$\tablenotemark{g}}
& \colhead{Z\tablenotemark{h}} & \colhead{${\rm L}_{48}$\tablenotemark{i}} \nl
\colhead{position} & \colhead{} & \colhead{} & \colhead{}& \colhead{} 
& \colhead{rad.} }
\startdata
J0009+73   & 55.2 $\pm$ 8.0 & 8.6  & 119.81 & 10.47 & 28  & ?          & -  & -     & - \nl
J0445+62   & 17.8 $\pm$ 5.0 & 4.0  & 146.89 & 10.65 & 36  & ?          & -  & -     & - \nl
J0500+59   & 17.1 $\pm$ 4.6 & 4.2  & 150.52 & 10.25 & 62  & ?          & -  & -     & - \nl
J0720+71   & 17.2 $\pm$ 2.4 & 8.7  & 143.90 & 27.94 & 25  & 0716+714   & 6  & $> 0.3$ & $> 0.029$ \nl
J0740+18   & 15.2 $\pm$ 4.2 & 4.2  & 201.35 & 18.77 & 110 & 0735+178   & 51 & 0.424 & 0.040 \nl
J0744+54   & 16.4 $\pm$ 2.9 & 6.8  & 162.98 & 29.32 & 44  & ?          & -  & -     & - \nl
J0808+49   & 14.9 $\pm$ 2.9 & 6.0  & 169.06 & 32.56 & 74  & 0804+499   & 5  & 1.43  & 1.5 \nl
J0811$-$07 & 26.2 $\pm$ 6.2 & 5.2  & 228.86 & 14.33 & 47  & ?          & -  & -     & - \nl
J0826+70   &  8.1 $\pm$ 2.0 & 4.7  & 144.19 & 33.33 & 74  & 0836+710   & 73 & 2.172 & 4.2 \nl
J0831+24   & 24.6 $\pm$ 6.0 & 5.2  & 200.23 & 31.89 & 56  & 0827+243   & 11 & 2.046 & 0.79 \nl
J1315$-$34 & 18.0 $\pm$ 3.3 & 6.4  & 308.46 & 28.14 & 39  & 1313$-$333 & 51 & 1.21  & 1.3 \nl
J1326$-$43 & 17.5 $\pm$ 3.5 & 5.7  & 309.56 & 18.90 & 41  & Cen A?     & -  & -     & -   \nl
J1626$-$24 & 25.0 $\pm$ 4.2 & 6.7  & 352.63 & 16.60 & 25  & 1622$-$253 & 33 & ?     & - \nl
J1629$-$28 & 18.5 $\pm$ 3.7 & 5.5  & 350.42 & 13.81 & 67  & ?          & -  & -     & - \nl
J1635$-$14 & 13.4 $\pm$ 3.6 & 4.2  & 2.57   & 21.68 & 28  & ?          & -  & -     & - \nl
J1719$-$04 & 18.8 $\pm$ 4.7 & 4.5  & 17.84  & 18.10 & 65  & ?          & -  & -     & - \nl
J1734$-$13 & 21.8 $\pm$ 4.2 & 5.7  & 12.07  & 10.38 & 33  & 1730$-$130 & 26 & 0.902 & 0.53 \nl
J1738+51   & 23.8 $\pm$ 4.1 & 7.2  & 79.09  & 31.98 & 36  & 1739+522   & 28 & 1.375 & 2.0 \nl
J1835+59   & 65.2 $\pm$ 5.1 & 17.9 & 88.77  & 25.09 & 9   & ?          & -  & -     & - \nl
\enddata
\tablenotetext{a}{Right ascension and declination (epoch 2000) of the point source in the 
format hhmm$\pm$dd.}
\tablenotetext{b}{Flux of photons with energy $>100$ MeV in units of $10^{-8}$ photons cm$^{-2}$
s$^{-1}$.  The uncertainty is $1 \sigma$, statistical only.}
\tablenotetext{c}{A measure of the statistical significance of the detection, as described
in the text.}
\tablenotetext{d}{Galactic longitude and latitude in degrees.}
\tablenotetext{e}{Radius, in arcminutes, of the 95\% confidence error circle.}
\tablenotetext{f}{The name of the radio source with which this gamma source can be identified.}
\tablenotetext{g}{Distance, in arcminutes, between the radio source position and the best position
of the gamma source.}
\tablenotetext{h}{Redshift of identified radio source, if known.}
\tablenotetext{i}{Gamma ray luminosity in the 100 MeV to 10 GeV range, in units of 
$10^{48}$ erg~s$^{-1}$, assuming isotropic emission, $q_0 = {1 \over 2}$, and 
$H_0 = 75$ km s$^{-1}$ Mpc$^{-1}$.
A $K$ correction factor of $(1+z)^{-(\alpha+2)}$ has been applied.}

\end{deluxetable}

\begin{deluxetable}{cccccc}
\tablecaption{Fluxes in different energy bands \label{table2}}
\tablehead{
\colhead{EGRET\tablenotemark{a}}
 & \colhead{${\rm F}_{30-100}$\tablenotemark{b}}
 & \colhead{${\rm F}_{100-300}$\tablenotemark{b}}
 & \colhead{${\rm F}_{300-1000}$\tablenotemark{b}}
 & \colhead{${\rm F}_{>1000}$\tablenotemark{b}}
 & \colhead{Spectral\tablenotemark{c}} \nl
\colhead{Position} & \colhead{} & \colhead{} & \colhead{} & \colhead{}
 & \colhead{Index} }
\startdata
J0009+73   & 49  $\pm$ 39 & 18.8 $\pm$ 6.3 & 17.8 $\pm$ 3.5 & 8.6 $\pm$ 2.4 & $-1.58 \pm 0.20$ \nl
J0445+62   & 43  $\pm$ 29 & 16.1 $\pm$ 4.8 & 2.5  $\pm$ 1.6 & 1.4 $\pm$ 0.8 & $-2.31 \pm 0.39$ \nl
J0500+59   & $ < 48$      & 6.5  $\pm$ 4.1 & 6.4  $\pm$ 1.8 & $ < 1.6$      & $-1.99 \pm 0.52$ \nl
J0720+71   & $ < 39$      & 12.0 $\pm$ 2.3 & 3.8  $\pm$ 0.9 & 1.2 $\pm$ 0.5 & $-2.12 \pm 0.24$ \nl
J0740+18   & 104 $\pm$ 28 & 13.3 $\pm$ 3.9 & $ < 4.0$       & $ < 1.3$      & $-3.54 \pm 0.83$ \nl
J0744+54   & $ < 31$      & 8.5  $\pm$ 2.5 & 4.9  $\pm$ 1.2 & 0.7 $\pm$ 0.5 & $-2.06 \pm 0.29$ \nl
J0808+49   & 71  $\pm$ 17 & 11.1 $\pm$ 2.6 & 3.0  $\pm$ 1.0 & $ < 0.6$      & $-2.72 \pm 0.38$ \nl
J0811$-$07 & 46  $\pm$ 35 & 21.5 $\pm$ 5.7 & 3.4  $\pm$ 2.1 & 1.7 $\pm$ 1.2 & $-2.35 \pm 0.55$ \nl
J0826+70   & 53  $\pm$ 13 & 8.6  $\pm$ 2.0 & 1.3  $\pm$ 0.6 & $ < 0.6$      & $-2.86 \pm 0.36$ \nl
J0831+24   & $< 63$       & 19.8 $\pm$ 5.5 & 3.9  $\pm$ 2.0 & 1.0 $\pm$ 1.2 & $-2.21 \pm 0.47$ \nl
J1315$-$34 & $ < 37$      & 7.6  $\pm$ 2.8 & 4.3  $\pm$ 1.2 & 1.4 $\pm$ 0.7 & $-1.84 \pm 0.30$ \nl
J1326$-$43 & 110 $\pm$ 24 & 15.5 $\pm$ 3.3 & 2.1  $\pm$ 1.1 & 0.7 $\pm$ 0.5 & $-2.85 \pm 0.38$ \nl
J1626$-$24 & 46  $\pm$ 22 & 12.2 $\pm$ 3.8 & 8.2  $\pm$ 1.6 & 0.5 $\pm$ 0.5 & $-2.27 \pm 0.24$ \nl
J1629$-$28 & $ < 41$      & 12.2 $\pm$ 3.5 & 3.7  $\pm$ 1.3 & 0.7 $\pm$ 0.5 & $-2.20 \pm 0.32$ \nl
J1635$-$14 & 39  $\pm$ 22 & 5.4  $\pm$ 3.3 & 3.4  $\pm$ 1.3 & 1.1 $\pm$ 0.6 & $-1.87 \pm 0.49$ \nl
J1719$-$04 & 112 $\pm$ 28 & 14.6 $\pm$ 4.5 & 5.5  $\pm$ 1.7 & $ < 1.5 $     & $-2.61 \pm 0.42$ \nl
J1734$-$13 & 60  $\pm$ 25 & 13.4 $\pm$ 3.9 & 6.3  $\pm$ 1.6 & $ < 1.5 $     & $-2.39 \pm 0.27$ \nl
J1738+51   & 34  $\pm$ 24 & 20.3 $\pm$ 3.9 & 3.4  $\pm$ 1.2 & 1.6 $\pm$ 0.9 & $-2.23 \pm 0.38$ \nl
J1835+59   & 70  $\pm$ 25 & 32.1 $\pm$ 4.1 & 20.3 $\pm$ 2.3 & 8.1 $\pm$ 1.5 & $-1.75 \pm 0.11$ \nl
\enddata
\tablenotetext{a}{Celestial position, as in Table 1.}
\tablenotetext{b}{Photon flux in four different energy bands, in units of $10^{-8}$
photons cm$^{-2}$ s$^{-1}$.  Uncertainties are $1 \sigma$, statistical only.
Upper limits are 95\% confidence.}
\tablenotetext{c}{Photon number index defined by ${dN \over dE} \propto E^\alpha$, based on fitting the fluxes in columns 2--5.
Uncertainties are derived from $\chi^2_{min} +2.3$.}
\end{deluxetable}

\begin{deluxetable}{cccc}
\tablecaption{Significant individual detections \label{table3}}
\tablehead{
\colhead{EGRET} & \colhead{date} & \colhead{$\rm F_{>100}$} &
 \colhead{spectral} \nl
\colhead{position} & \colhead{} & \colhead{} & \colhead{index} }
\startdata
J0009+73 & 1992 Jul 16 -- Aug 6  &  72.7 $\pm$ 13.4 & $-1.58 \pm$ 0.24 \nl
         & 1993 Feb 25 -- Mar 9  &  36.7 $\pm$  9.9 & $-1.47 \pm$ 0.47 \nl
J0720+71 & 1992 Jan 10--23       &  22.0 $\pm$  4.7 & $-2.31 \pm$ 0.33 \nl
         & 1992 Mar  5--19       &  52.4 $\pm$ 13.2 & $-2.57 \pm$ 0.49 \nl
         & 1993 Jul 13--27       &  18.7 $\pm$  5.2 & $-1.63 \pm$ 0.47 \nl
J0744+54 & 1993 Jun 29 -- Jul 13 &  38.1 $\pm$  7.7 & $-1.96 \pm$ 0.29 \nl
         & 1993 Jul 13--27       &  17.8 $\pm$  5.4 & $-1.78 \pm$ 0.65 \nl
J0826+70 & 1992 Jan 10--23       &  14.1 $\pm$  3.9 & $-2.62 \pm$ 0.36 \nl
         & 1992 Mar  5--19       &  37.6 $\pm$ 10.8 & $-2.47 \pm$ 0.49 \nl
J1835+59 & 1991 May 30 -- Jun 8  &  82.5 $\pm$ 15.4 & $-1.96 \pm$ 0.32 \nl
         & 1991 Sep 12--19       &  43.2 $\pm$ 13.9 & $-1.52 \pm$ 0.47 \nl
         & 1992 Mar  5--19       &  42.8 $\pm$ 13.0 & $-2.22 \pm$ 0.53 \nl
         & 1992 Nov 17--24       &  92.5 $\pm$ 26.0 & $-2.04 \pm$ 0.53 \nl
         & 1992 Nov 24 -- Dec 1  &  94.4 $\pm$ 21.6 & $-1.88 \pm$ 0.39 \nl
         & 1992 Dec  1--22       &  92.2 $\pm$ 13.7 & $-1.91 \pm$ 0.24 \nl
         & 1993 Mar  9--23       &  54.0 $\pm$  8.1 & $-1.46 \pm$ 0.21 \nl
\enddata
\end{deluxetable}

\clearpage
\section*{Figure Captions}
\figcaption{Average spectra of the nineteen detected point sources with
the best-fitting power-law models.
The dotted lines represent $2 \sigma$ upper limits. \label{fig1}}
\figcaption{Light curves (E $>$ 100 MeV) for three of the variable sources.
Each data point represents an entire CGRO viewing period. \label{fig2}}
\figcaption{Energy spectra of J1835+59 in individual viewing periods,
with the best-fitting power-law models.
The dotted lines represent $2\sigma$ upper limits. \label{fig3}}


\clearpage
\begin{references}
\reference{} Angel, J. R. P., \& Stockman, H. S. 1980, \araa, 18, 321
\reference{} Beall, J. H., \& Rose, W. K. 1980, ApJ, 238, 539
\reference{} Becker, R. H., White, R. L., \& Edwards, A. L. 1991,
ApJS, 75, 1
\reference{} Bertsch, D. L. et al. 1989, in Proc. Gamma-Ray Observatory
Science Workshop, ed. W. N. Johnson (Greenbelt MD: NASA), 4-49
\reference{} Bertsch, D. L., Dame, T. M., Fichtel, C. E., Hunter, S. D.,
Sreekumar, P., Stacy, J. G., \& Thaddeus, P. 1993, ApJ, 416, 587
\reference{} Bertsch, D. L., et al. 1995, in preparation
\reference{} Burbidge, G., \& Hewitt, A. 1992, in Variability of Blazars,
ed. E. Valtaoja \& M. Valtonen (Cambridge: Cambridge), 4
\reference{} Collmar, W., et al. 1993, in Compton Gamma-Ray Observatory,
AIP Conf. Proc. 280,
ed. M. Friedlander, N. Gehrels, \& D. J. Macomb (New York: AIP), 483
\reference{} Dingus, B. L., et al. 1995, ApJ, submitted
\reference{} Feigelson, E. D., Schreier, E. J., Delvaille, J. P.,
Giacconi, R., Grindlay, J. E., \& Lightman, A. P. 1981, ApJ, 251, 31
\reference{} Fichtel, C. E., et al. 1994a, ApJS, 94, 551
\reference{} Fichtel, C. E., et al. 1994b, in The Second Compton Symposium,
AIP Conf. Proc. 304, ed. C. E. Fichtel, N. Gehrels, \& J. P. Norris
(New York: AIP), 721
\reference{} Grindlay, J. E. 1975, ApJ, 199, 499
\reference{} Helou, G., Madore, B. F., Schmitz, M., Bicay, M. D., Wu, X.,
\& Bennett, J. 1991, in Databases and On-Line Data in Astronomy,
ed. D. Egret \& M. Albrecht (Dordrecht: Kluwer), 89
\reference{} Hewitt, A., \& Burbidge, G. 1993, ApJS, 87, 451.
\reference{} Hui, X., Ford, H. C., Ciardullo, R., \& Jacoby, G. H. 1993,
ApJ, 414, 463
\reference{} Hunter, S. D., Digel, S. W., de Geus, E. J., \& Kanbach, G.
1994, ApJ, 436, 216
\reference{} Hunter, S. D., et al. 1995, in preparation
\reference{} Impey, C. D., \& Tapia, S. 1990, ApJ, 354, 124
\reference{} de Jager, O. C., \& Harding, A. K. 1992, ApJ, 396, 161
\reference{} Kanbach, G., et al. 1988, \ssr, 49, 69
\reference{} Kanbach, G., et al. 1989, in Proc. Gamma-Ray Observatory
Science Workshop, ed. W. N. Johnson (Greenbelt MD: NASA), 2-1
\reference{} Kinzer, R. L., et al. 1995, ApJ, 449, 105
\reference{} Lampton, M., Margon, B., \& Bowyer, S. 1976, ApJ, 208, 177
\reference{} Lin, Y. C., et al. 1995a, ApJ, 442, 96
\reference{} Lin, Y. C., et al. 1995b, ApJ, submitted
\reference{} Mattox, J. R., et al. 1996, ApJ, submitted
\reference{} Meier, D. L., et al. 1989, AJ, 98, 27
\reference{} Michelson, P. F., et al. 1994, in The Second Compton Symposium,
AIP Conf. Proc. 304, ed. C. E. Fichtel, N. Gehrels, \& J. P. Norris
(New York: AIP), 602
\reference{} Milne, D. K. 1979, Austral. J. Phys., 32, 83
\reference{} von Montigny, C., et al. 1995, ApJ, 440, 525
\reference{} Mukherjee, R., et al. 1995, ApJ, 445, 189
\reference{} Mushotzky, R. F., Serlemitsos, P. J., Becker, R. H.,
Boldt, E. A., \& Holt, S. S. 1978, ApJ, 220, 790
\reference{} Nice, D. J., Sayer, R. W., \& Taylor, J. H. 1994, in The Second Compton Symposium,
AIP Conf. Proc. 304, ed. C. E. Fichtel, N. Gehrels, \& J. P. Norris
(New York: AIP), 86
\reference{} Nolan, P. L., et al. 1994, in The Second Compton Symposium,
AIP Conf. Proc. 304, ed. C. E. Fichtel, N. Gehrels, \& J. P. Norris
(New York: AIP), 360
\reference{} Quinn, J., et al. 1995, IAU Circular 6178
\reference{} Seward, F. 1990, ApJS, 73, 781
\reference{} Sieber, W., Salter, C. J., \& Mayer, C. J. 1981, A\&A, 103, 393
\reference{} Skibo, J. G., Dermer, C. D., \& Kinzer, R. L. 1994, ApJ, 426, L23
\reference{} Sreekumar, P. et al. 1995, ApJ, submitted
\reference{} Steidel, C. C., \& Sargent, W. L. W. 1991, ApJ, 382, 433
\reference{} Stevens, J. A., Litchfield, S. J., Robson, E. I., Hughes,
D. H., Gear, W. K., Ter\"asranta, H., Valtaoja, E., \& Tornikoski, M.
1994, ApJ, 437, 91
\reference{} Sturner, S. J., \& Dermer, C. D. 1995, A\&A, 293, L17
\reference{} Thompson, D. J., et al. 1993a, ApJS, 86, 629
\reference{} Thompson, D. J., et al. 1993b, ApJ, 415, L13
\reference{} Thompson, D. J., et al. 1994, ApJ, 436, 229
\reference{} Thompson, D. J., et al. 1995, ApJS, in press
\reference{} Witzel, A., Schalinski, C. J., Biermann, P. L., Krichbaum, T. P.,
\& Johnston, K. J. 1988, A\&A, 206, 245
\reference{} Zdziarski, A. A., Ghisellini, G., George, I. M., Svensson, R.,
\& Fabian, A. C. 1990, ApJ, 363, L1
\end{references}
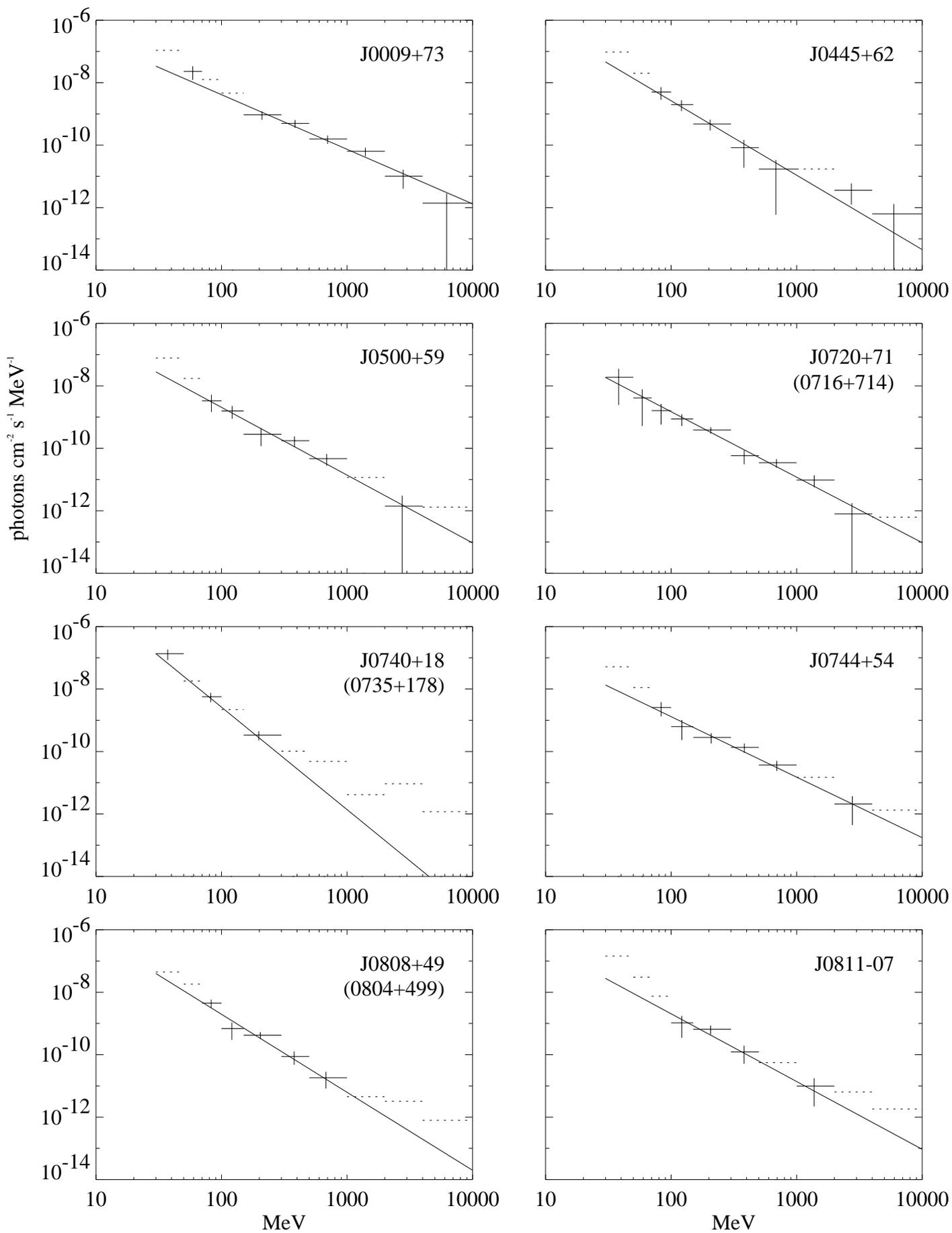
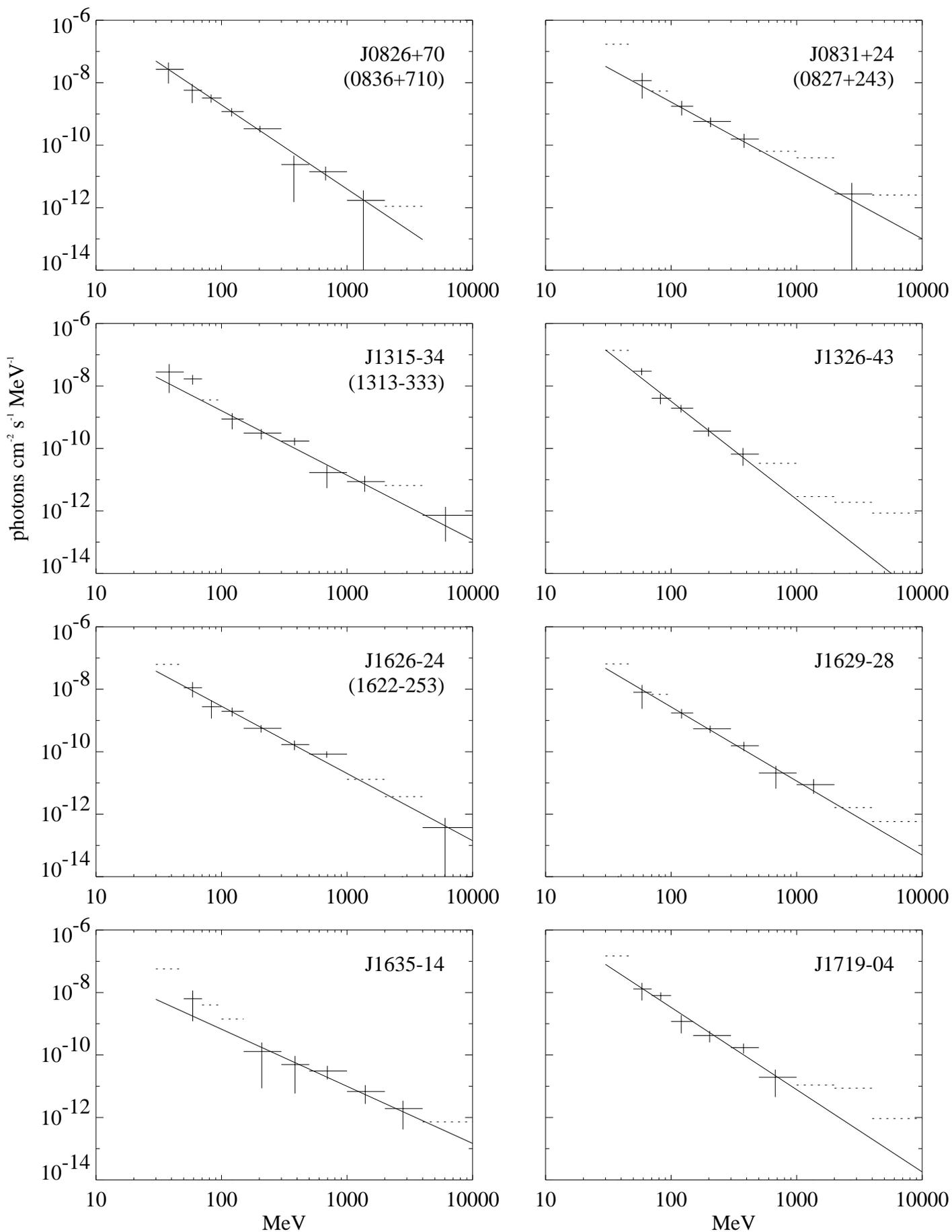
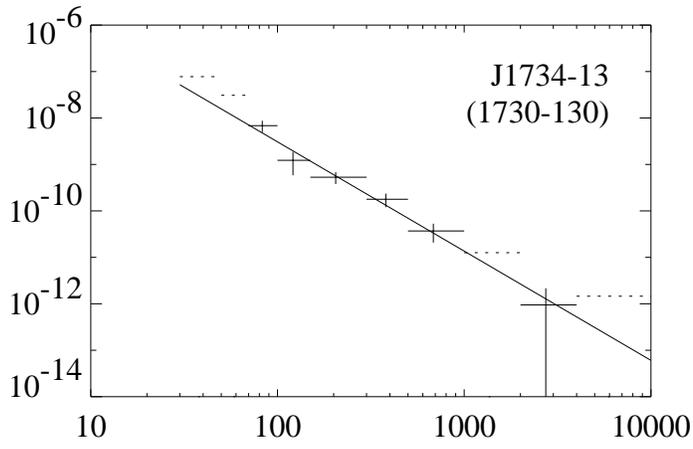
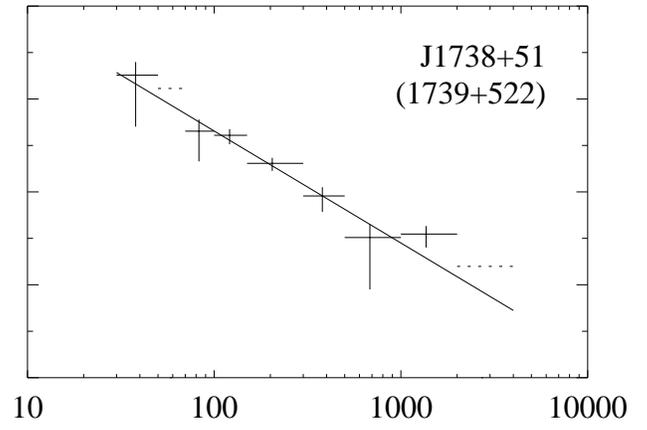
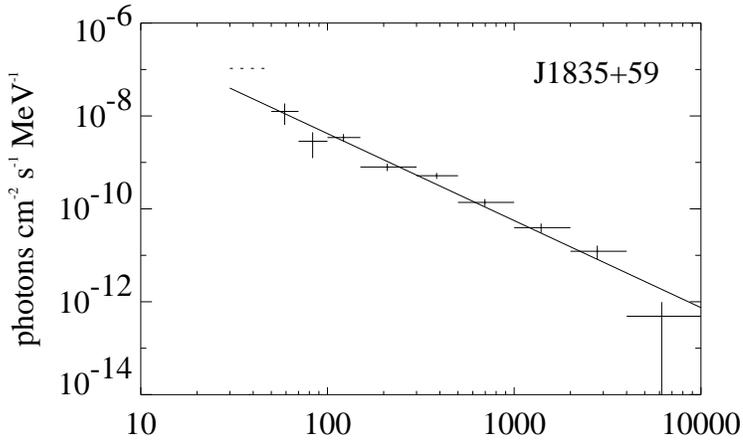
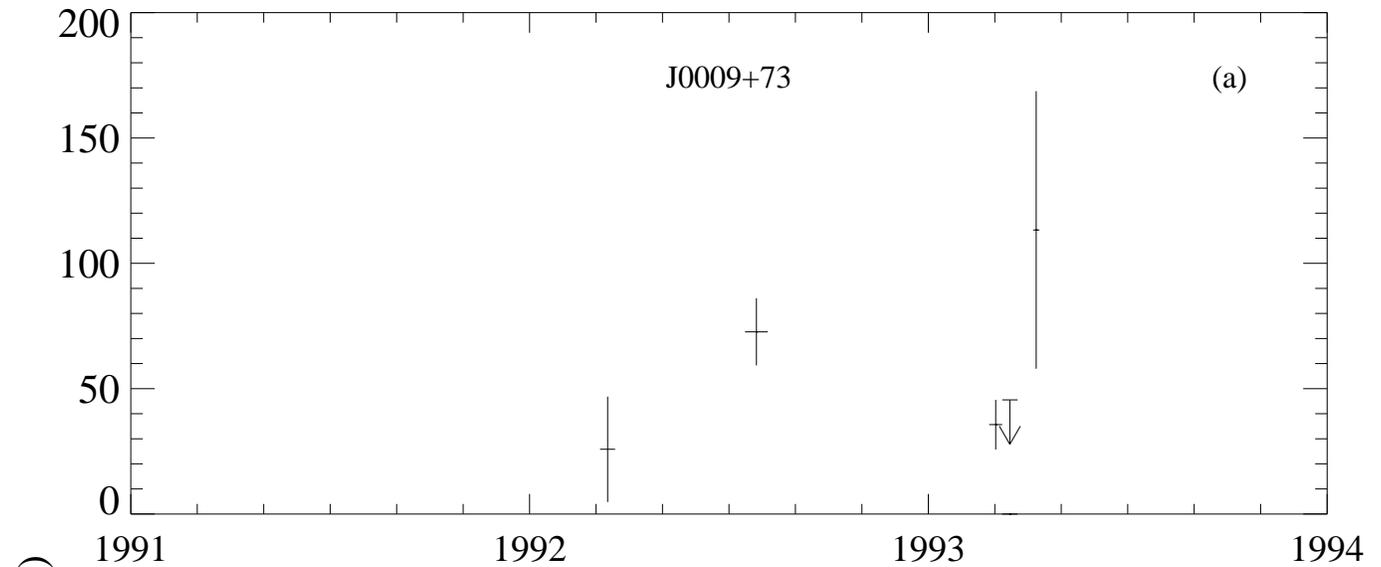
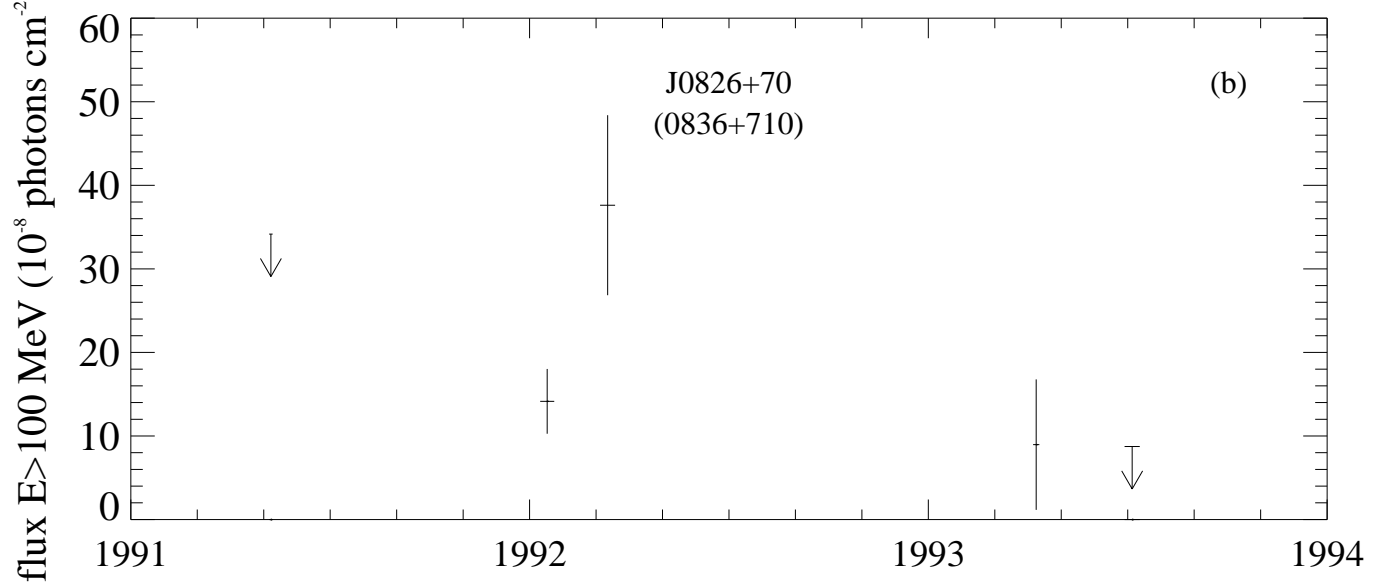
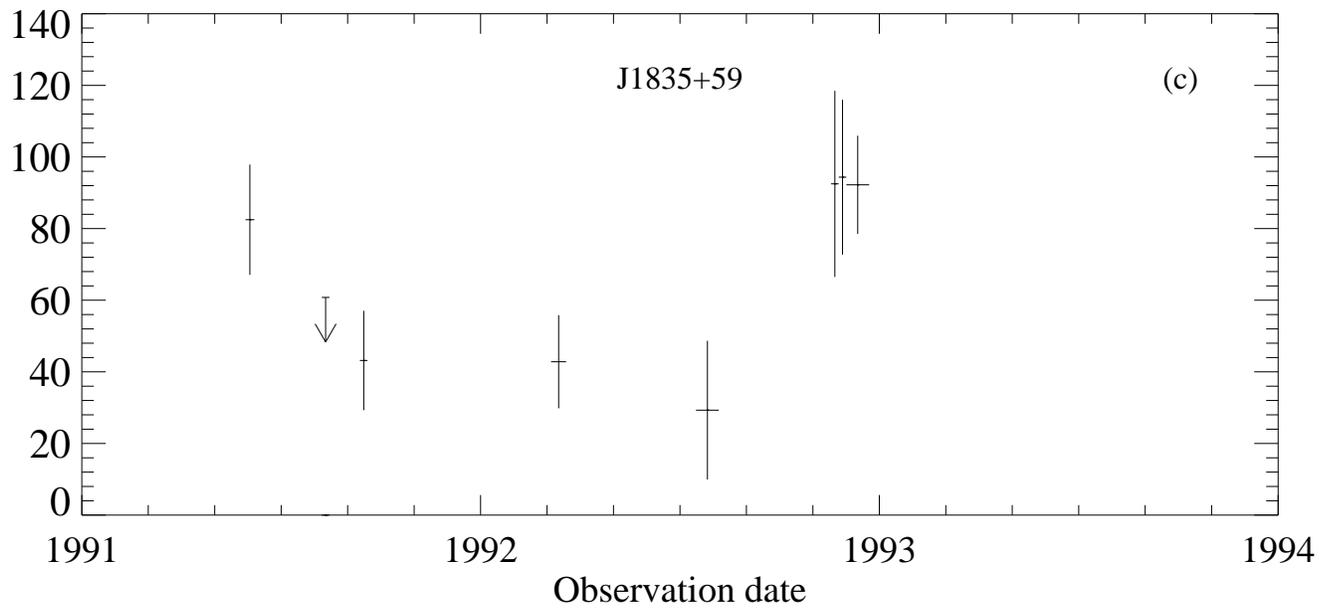
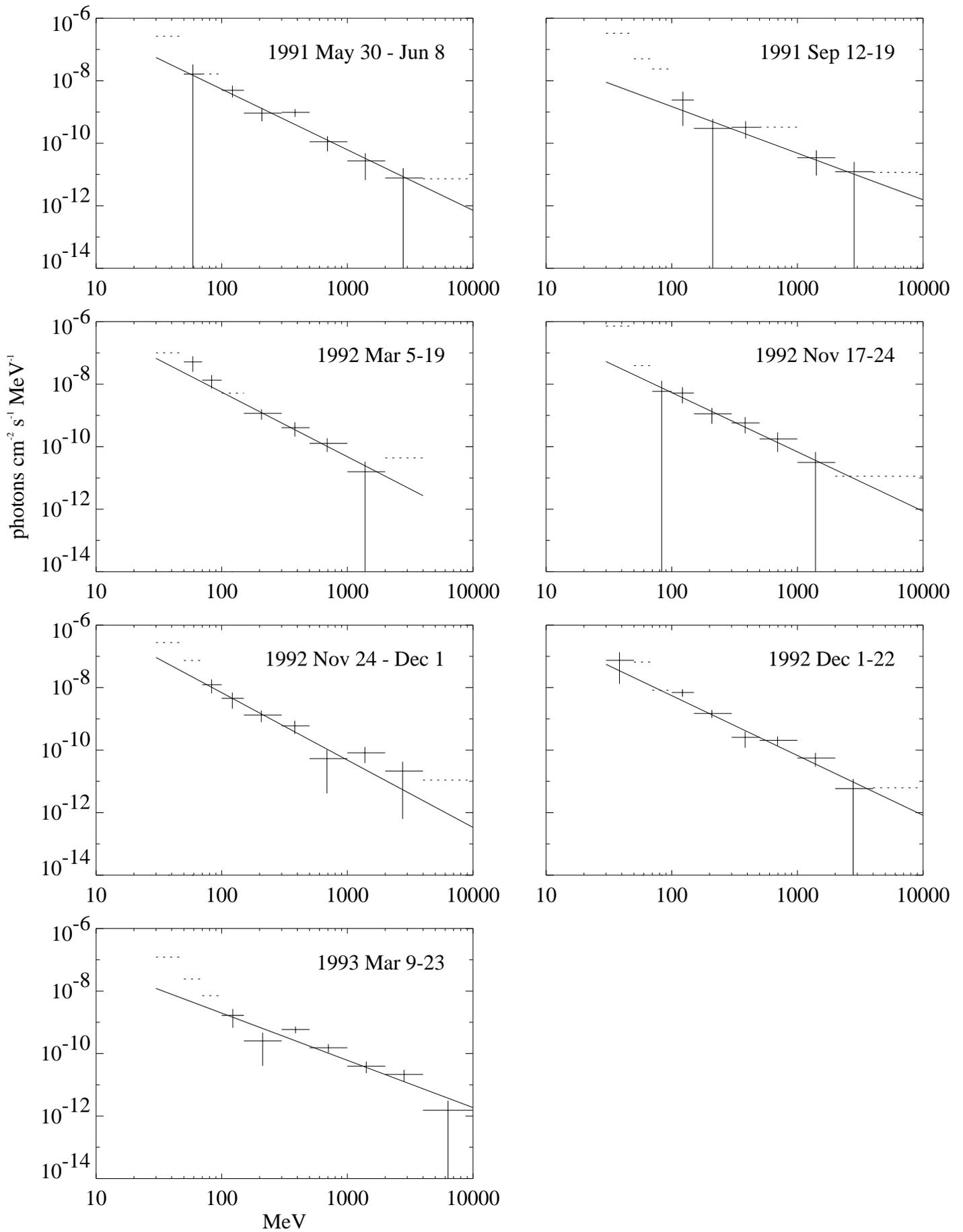
\end{document}